\documentclass[12pt]{article}
\usepackage{amssymb}
\usepackage{graphicx}
\usepackage{enumerate}

\usepackage{lipsum}

\newcommand\blfootnote[1]{%
  \begingroup
  \renewcommand\thefootnote{}\footnote{#1}%
  \addtocounter{footnote}{-1}%
  \endgroup
}

\begin{document}
\bibliographystyle{plain}
\setlength{\baselineskip}{1.5\baselineskip}
\newtheorem{theorem}{Theorem}
\newtheorem{lemma}{Lemma}
\newtheorem{corollary}{Corollary}
\newcommand{\eqref}[1]{(\ref{#1})}
\newcommand{\boldm}[1]{\mbox{\boldmath$#1$}}
\newcommand{\bigO}[1]{{\mathcal{O}}\left( {#1 }\right)}
\newcommand{\bigOp}[1]{{\mathcal{O}}_p\left( {#1} \right)}
\newcommand{\smallO}[1]{{o}\left( {#1} \right)}
\newcommand{\sign}[1]{{sign}\left( {#1} \right)}
\newcommand{\smallOp}[1]{{o}_p\left( {#1} \right)}
\newcommand{\half}{\small \frac{1}{2}}
\newcommand{\fourth}{\small \frac{1}{4}}
\renewcommand{\baselinestretch}{1}
\newenvironment{proof}[1][Proof]{\noindent\textbf{#1.} }{\ \rule{0.5em}{0.5em}}

\title{Canonical Regression Quantiles \\
with application to CEO compensation \\
and predicting company performance} 

\author{Stephen Portnoy{$ ^1 $} and Joseph Haimberg{$ ^2 $} }

\date{\today}

\maketitle 

\begin{abstract}
In using multiple regression methods for prediction, one often considers the linear combination
of explanatory variables as an index. Seeking a single such index when here are multiple responses 
is rather more complicated. One classical approach is to use the coefficients from the leading
canonical correlation. However, methods based on variances are unable to disaggregate
responses by quantile effects, lack robustness, and rely on normal assumptions for inference.
We develop here an alternative regression quantile approach and apply it to an empirical study of
the performance of large publicly held companies and CEO compensation. The initial results
are very promising.
\end{abstract}

\blfootnote{
$\qquad \,$ company performance, canonical correlation\\
$\quad$}

\footnotetext[1]{\noindent 
Professor, Department of Statistics, University
of Illinois at Urbana-Champaign \\
\smallskip
$\quad$ corresponding email: sportnoy@illinois.edu \\ }
\footnotetext[2]{\noindent 
Portland State University, Mark O. Hatfield School of Government}

\newpage

\section{Introduction.}

We develop a novel approach for using the explanatory $X$-variables to find an index
that predicts future outcomes given by $Y$-variables in a quantile regression setting. The
approach is motivated by a specific financial data set listing various measures of the
behavior and performance of some large U.S. companies. The aim is to
use prior company data to predict future performance, and seeks to define 
an index (in terms of past data) with better predictive power than CEO total
compensation for regression quantile methods.  A classical
approach might be to use the  leading Canonical Correlation,
which provides a  linear combination of $X$-variables that is most highly correlated with
some linear combination of the $Y$-variables.  However, the inability of correlation methods
to disaggregate responses by quantile effects, their inherent lack of robustness, and their need
for normal assumptions strongly motivates the search for an alternative
regression quantile approach.

The empirical data set concerns research into the relationship between CEO compensation
and corporate performance. Much previous financial analysis had failed to yield meaningful
results and suggests decoupling of CEO compensation from corporate performance.
Recent work by Barton et al (2017) suggested a predictive model where public
corporations that adopt operational and financial long-termism will achieve superior
financial and market results. The suggested predictive power of long-termism
could provide a quantitative measure of future corporate performance that are tied to current
CEO’s actions. If true, such models could be used to set current CEO compensation. Corporate
ownership structure can influence CEO compensation to reflect
ownership powers rather than corporate performance. Therefore, we limited our research
to non-controlled public corporations – those with a single class of shares and without
blockholders (defined as stock ownership exceeding 10\%). 

The analysis here develops an index with noticeably better predictive properties than
CEO compensation. We focus on predicting two years into the future based on 5 years
of past data (which accords with a long-term view). While the results are preliminary, they
strongly support the potential value of the ``canonical'' regression quantile approach. 
A more complete analysis of the data together with a careful consideration of the financial
issues appears in Haimberg and Portnoy (2020).

\section{Data.}

The study assembled a database by extracting financial information on publicly 
traded companies on the S\&P500 index with continuous operations from 2009 to 2018.
 Data sources include SEC Form DEF 14A (also refer to as Proxy Statement) and 
 SEC 10K (Annual Report) that are filed annually with the Securities and
 Exchange Commission (SEC). 

Our data includes only companies without any blockholders 
and to companies with a single class of shareholders to prevent potential bias due to 
controlling shareholders. We also excluded companies that either went through a
merger or new companies that have been in existence for less than 10 years to remove
longitudinal data gaps.  We identified 100 companies satisfying these criteria.

In a attempt to model long-termism, Barton (2017) suggested a set of independent variables
to predict the dependent variables that measure future performance. We used the same set
of variables with the exception of one of the independent variables that measure external input
based on analysts’ earnings target. We restrict our data to the period from 2009 to 2018 in order
to analyze a rather stable economic environment without any major financial shocks.

The explanatory $X$-variables and performance response $Y$-variables are
defined in the Tables 1 and 2. To include technical definitions, ``Accruals'' means
Operating Net Income minus Operating Cash Flow, 
``True Earnings'' means  Operating Cash Flow divided by  Diluted Number of Shares, and
``opportunity cost'' means the Total Invested Capital times the Weighted Average Cost of Capital,
and ``Shares'' means Diluted Number of Shares.

\begin{table}[htb] \centering 
  \caption{Explanatory Variables} 
  \label{EV} 
\footnotesize 

\begin{tabular}{@{\extracolsep{5pt}}lll} 
\hline 
variable  &  abbr.  &  definition
 \\ 
\hline \\
Investment Ratio & IR &Ratio of capital expenditures to Depreciation \\
Earning Quality Ratio &EQ & Accruals as share of the revenues \\
Margin Growth Ratio & MG & Earning growth divided by revenue growth \\
Earnings per Share	& EPS & EPS Growth less True Earnings Growth \\
CEO Total Package	& CEOtot &	Salary, bonus, stock, stock options and other pay \\
\hline \\
\normalsize 
\end{tabular} 
\end{table} 

\begin{table}[htb] \centering 
  \caption{Response Variables} 
  \label{RV} 
\footnotesize 

\begin{tabular}{@{\extracolsep{5pt}}lll} 
\hline 
variable  &  abbr.  &  definition
 \\ 
\hline \\
Revenues & REV & Gross sales – returns – discounts revenues \\
Earnings & Earn & Earnings available to shareholders \\
Economic Profits & Eprof & Net Income – opportunity cost \\
Market Capitalization	 & MCap & Share Price at year-end times Shares \\
Total Shareholder Return & TSR &  Capital gain on shares plus dividends \\
\hline \\
\normalsize 
\end{tabular} 
\end{table} 

\bigskip

As described below, the CEOtot variable is used both in developing the predictive index 
and compared to the index in predicting future performance. In addition, there is a
categorical variable with 6 levels giving the type of company:  industrial, health, consumer,
energy, tech, and utility. See Haimberg and Portnoy (2021) for further discussion of the
issues involved in analyzing this data,

\section{Statistical Methodology.} \label{sec:method}

The method of Canonical Regression Quantiles seeks to find a linear combination of the response
variables that is best predicted by a linear combination of the explanatory variables in terms of 
regression quantile objective functions.
The basic idea is as follows: let $\rho_\tau(u) = u ( \tau - I(u<0))$ be the quantile objective function
so that the $\tau$th quantile, $Q_\tau$ of a sample $\{x_i\}$ satisfies
\begin{equation} \label{Qdeff}
Q_\tau = \min_\theta \, \sum_i \, \rho_\tau(x_i - \theta)
\end{equation}

Given a data matrix, $X$, of explanatory variables (generally including a constant intercept column)
and a data matrix $Y$ of response variables,
we would like to define the first canonical regression quantile as the pair of coefficient vectors,
$(\alpha, \, \beta)$ achieving
\begin{equation} \label{canobj}
\min_{\alpha,\beta} \, \sum_i \, \rho_\tau( x_i' \beta \, - \, y_i' \alpha ) 
\end{equation}
where $x_i$ and $y_i$ denote the $i$th rows of $X$ and $Y$ respectively. Unfortunately,
this won't define unique solutions since the objective function in \eqref{canobj} is invariant
under scale changes. In the case of canonical correlations, this problem is resolved by imposing 
orthogonality (or, more generally, a quadratic norm identity). However, to reflect the quantile
setting and to avoid the lack of robustness of quadratic measures, it would be preferable to
impose the following $L_1$ identity:
\begin{equation} \label{side0}
\sum_j \, | \alpha_j | = 1
\end{equation}
where $\alpha_j$ are the coordinates of $\alpha$. 
Minimizing \eqref{canobj} subject to \eqref{side0} can be expressed as a linear
programming problem, and one could stop here. However, settings like the application here 
suggest the need for an additional element.

In many (most) applications, the response variables measure related outcomes, and so we
expect them to be positively correlated and to be monotonic in the predictors. 
This suggests imposing the side conditions:
\begin{equation} \label{side0}
\sum_j \,  \alpha_j  \, = \, 1 \qquad \alpha_j  \geq 0 \quad j = 1 , \cdots , q 
\end{equation}
Minimizing \eqref{canobj} subject to this constraint is a constrained regression quantile problem.
The solution to \eqref{canobj} is well defined,
and easily computable since the linear constraints perfectly match the linear
programming form of the regression quantile problem (see Koenker (2005), p. 213).
Thus, letting $(\hat\alpha , \, \hat\beta)$ solve \eqref{canobj} subject to \eqref{side0}, 
we consider $\hat\alpha$ and $\hat\beta$ to define predictive and response indices (respectively),
and call $\, x_i' \beta \,$ and $\, y_i' \alpha \,$ the observed predictive and response 
indices corresponding to the $i$-th observation (that is, the $i$-th row of $X$ and $Y$).

If $\, \alpha_j > 0 \,$ (for all $j$), it is possible in theory to reduce to a standard regression quantile problem.
From the side condition \eqref{side0}, we can write $\, \alpha_1 = 1 - \sum_{j=2}^q \alpha_j \,$
and substitute into the objective function \eqref{canobj} to get
$$
\min_{\alpha,\beta} \, \sum \left( \rho_\tau( y_{i,1} \, - \sum_{j=2}^q y_{i,j} \alpha_j  \, + \,  x_i' \beta \right) \,\, .
$$
The solution to this minimization problem provides $\hat\beta$ and
$(\hat\alpha_j \, : \,\, j = 2 , \, \cdots \, , \, q ) $ (with $\, \alpha_1 = 1 - \sum_{j=2}^q \alpha_j \,$) as
standard regression quantile coefficients. Furthermore, if all $ \alpha_j > 0 \,$, the estimates will all
be positive with probability tending to 1 asymptotically. The coefficient estimates
will satisfy all known results for regression quantiles
(finite sample, asymptotic, and computational; see Koenker (2005)).

Unfortunately, the presence of
constraints can seriously affect theoretical properties if some ${\hat\alpha}_j$ are negative
or are on the boundary. This can occur with positive probability for finite samples even when
all $\, \alpha_j \,$ are positive. In such cases, the constrained solution will be a projection
into the constraint set, and the asymptotic theory  becomes rather complicated.
See Chernoff (1954) for an early description of such asymptotics. 

Specifically, if the $\alpha$-constraints fail, the estimates will lie on the boundary with  
non-vanishing probability (asymptotically). In this case, Andrews (2000) showed that standard
bootstrap methods fail, but that a subsample bootstrap whose subsample size tends to infinity 
more slowly than $n$ will provide appropriate inferences for certain constrained maximum likelihood
estimators. Later work extended these results in various directions, and the Andrews bootstrap
should work here.

Thus, the focus here will use a weighted version of the Andrews' bootstrap that is informed
by recent research in this area. One complication is that the
rank of the $X$-matrix is not a small fraction of the sample size, $n$. Thus, repeated
observations in bootstrap samples can lead to singular (or nearly singular) design matrices. To
avoid this problem, weighted resampling methods will be  used here. Specifically, a weighted
version of a subsample bootstrap of size $\, m \,$ (proportional to $\log{n}$) will be used here.
A subsample of size $m$ will be given independent negative exponential weights, and the remaining 
observations will be given $(1/n)$ times independent uniform weights; and this subsampling process
will be repeated $R$ times to provide a ``bootstrap'' sample. In theory, this should be asymptotically
equivalent to Andrews' bootstrap procedure, and this was used for all results reported here.

The problem of estimates potentially falling on the boundary might be addressed by an
alternative approach.  Cavaliere and Georgiev (2019)
suggest that the main problem generating bootstrap inconsistency in such cases occurs when the
probability that a parameter falls on the bounder differs substantially between the
bootstrap distribution and the distribution of the estimator under the model. To keep the difference
between the bootstrap and empirical probabilities small, a weighted version of a delete-$\sqrt{n}$
jackknife might be used successfully. Portnoy (2013) provides evidence that such a 
procedure is appropriate for censored regression quantiles, where it is also important to keep the
resampling distribution close to the empirical distribution. This method was tried for most of the
runs described below. It provided roughly similar estimates of standard deviations, though often
somewhat smaller. We also computed standard errors for predictions using CEOtot and the Index
using the bootstrap method internal to {\tt rq}. That is, we computed standard errors conditional
on the Index, ignoring variability in the data used to construct the index. The observed differences
were sufficiently small that our conclusions would not be affected.

\bigskip

\newpage

\section{Results of the data analysis}.

For the data described above,  a ``long-term" perspective  suggests using several years of data to
predict a small number of years ahead. Given only 10 years of data, we focus on
using 5 years data to predict 2 years ahead. Past values of both X and Y variables will be
used, perhaps suggestive of ``Granger Causality" (Granger, 1980). In addition, we will include  
CEO total compensation (CEOtot) in the $X$ variables to define the index and also as a predictor
for the response two years ahead. Assuming we have 
observed 7 years, we use the canonical regression quantile method above and, as an alternative,
classical canonical correlations to predict the 5 Y-variables in the current year. As described in detail
below, we use 
the $\beta$-coefficients for the first 5 years data to produce a prediction index; and then use
this index to predict 2 years into the future. 

With only 100 companies, it is not reasonable to use all the $X$ and $Y$ data over 5 years.
From the long-term perspective, we may consider replacing each variable by simple 
aggregates of the 5 years of data. Specifically, we consider two such aggregates:
a discounted average, discounting 5\% each year into the past (essentially an exponential smoothing);
and the minimum of the 4 differences between successive years. The first measures overall 
performance while the second measures some form of stability. This leads to 2 measurements
for each of the 4 $X$ variables, the 5 response $Y$ variables, and CEOtot:  a total of 20 variables.
In addition, we introduce an 
indicator variable for each of the 6 industry types (implicitly including an intercept). This creates
a  100 by 26 matrix of explanatory variables (for each time period), that we denote by $X^*$. 
Letting $Y^*$ denote the response variables 2 years ahead of the data for $X^*$ 
(that is, two years ahead of the ``current" year), we apply 
two ``canonical" analyses to develop predictor ($X$) and response ($Y$) indices. The first is
the canonical regression quantile method described above and the second used the vectors
corresponding to the leading (classical) canonical correlation.
We then use these indices applied to the data from the current year and the previous 4 years
to predict each of the 5 $Y$-values two years ahead.  To be consistent with the basic formulations, we
use Least Squares to predict 2 years ahead for classical canonical correlation index, and we
use median ($L_1$) regression to predict for the canonical regression quantile index here.
We then compare the predictions with the observed $Y$-values (2 years ahead).

To be precise, for example, we use the $X$ and $Y$ data from years 2009 to 2013 to 
predict $Y$ data in 2015
in order to create the index coefficients, and then apply these coefficients to data from years
2011 to 2015 to predict $Y$ variables in year 2017 (by a regression quantile estimator with 
proportion $\, \tau = .5$). We then repeat this process for data from 2010 to 2014 to create an 
index for predicting $Y$ variables in year 2018. Of course, other spans of years can be used, both
for creating the predictor indices and for predicting into the future. As remarked after presenting 
the statistical results below, runs using other choices were done. Though the details of the alternative
analyses differed, there were no substantive differences in conclusions.

As a final remark concerning the data set-up, note that, like typical economic measurements,
 the $Y$ variables were expected to be generally positive values but with clear non-normal
 variability. Thus, we followed common economic
practice and took logs. Unfortunately, some values were actually negative, especially for the
variables ``Economic Profits" and ``Total Shareholder Returns". Based on the rather large
variability in the negative values for these variables, 
we decided to use a signed log-transformation: sign(y)$\, \log | y | \,$  Some runs with
the more usual $\, \log (\max(1, \, y) ) \,$ transformation
were done, and they showed almost no difference (often in the third significant figure or smaller). 

To analyze the results, we consider three predictors: the index predictors using each
of the two canonical methods and also predictions based on total CEO compensation,
CEOtot. To allow consistent
comparisons, we use both Least Squares and Median regression to predict performance
(2 years ahead) based on CEOtot. To elucidate the canonical regression quantile method,
we will look at the $\alpha$ and $\beta$ coefficients and their standard errors based on the
weighted version of Andrews' bootstrap  described above.
Numerical comparisons were made in terms of root-mean-squared-error (RMSE) and the 
mean-absolute-error (MAE) combining results for years 2017 and 2018. However, only
the MAE results are reported here, as MAE is more robust and less sensitive to non-normality. We
assess statistical variation in these values using the same weighted version of the Andrews' bootstrap. 
We also look at some graphical presentations to get a more intuitive idea of the
issues, focussing on the canonical regression quantile and the CEOtot predictors. Among the
plots considered, we present here
scatter plots of predictor index (from the canonical regression quantile approach) and
CEOtot versus the canonical regression quantile response index 2 years ahead. We also
present scatter plots of the index and CEOtot versus each of the response variables
separately, and plots of $y$ vs. $\hat{y}$ in terms of Q-Q plots. Finally, we provide
an alternative comparison of predictability of
future economic performance by presenting median regressions based on using
the current Index and CEOtot together to predict the response variables 2 years ahead.

All computer work was done in R (R core team, 2015). Median regression used the function {\tt rq}
from the R-package, {\tt quantreg}, while Least Squares regression used the base function {\tt lm}.
The results are as follows.

The $\alpha$ coefficients for the ``best'' predicted combination of response variables given in Table 3:
 
 \begin{table}[htb] \centering 
  \caption{$\alpha$ coefficients for Index} 
  \label{alpha} 
\footnotesize 

\begin{tabular}{@{\extracolsep{5pt}} c c c c c c } 
\hline 
year & logRev  &  logEarn & logEprof  &  logMarCap & logTSR
 \\ 
\hline \\
2017 & .884 & .003 & 0 & .113 & 0 \\
2918 & .998 & 0 & 0 & 0 & .002 \\
\hline \\
\normalsize 
\end{tabular} 
\end{table} 

Clearly, $\alpha$ is controlled almost entirely by log Revenue, and it would
likely be adequate to base $\beta$ on the regression of log Revenue alone on $X^*$. This was
done as a check on the canonical regression quantile approach, since standard bootstrap
inference for conditional quantile analysis is legitimate. As expected, the differences between
using log Revenue alone and the canonical regression quantile index were very small, often
in the third significant figure or smaller.

The $\beta$ coefficients and their t-statistics (using the weighted bootstrap standard errors) are in Table 4:

\newpage
 
\begin{table}  \centering 
  \caption{Beta coefficients for Index} 
  \label{betas} 
  \footnotesize
\begin{tabular}{@{\extracolsep{5pt}} ccccc} 
\\[-1.8ex]\hline 
\hline \\[-1.8ex] 
 & 2017 & t-stat & 2018 & t-stat \\ 
\hline \\[-1.8ex] 
intercept & $0.754$ & $1.529$ & $0.022$ & $0.046$ \\ 
indust & $0.113$ & $0.886$ & $0.217$ & $2.110$ \\ 
health & $0.280$ & $1.888$ & $0.320$ & $2.853$ \\ 
consum & $0.097$ & $0.651$ & $0.129$ & $1.103$ \\ 
energy & $0.277$ & $1.402$ & $0.402$ & $2.086$ \\ 
tech & $0.264$ & $1.678$ & $0.321$ & $2.240$ \\ 
IRwt & $$-$0.026$ & $$-$0.830$ & $$-$0.020$ & $$-$0.739$ \\ 
IRminD & $$-$0.013$ & $$-$0.236$ & $$-$0.024$ & $$-$0.421$ \\ 
EQwt & $$-$0.003$ & $$-$0.755$ & $0.002$ & $0.408$ \\ 
EQminD & $$-$0.001$ & $$-$0.266$ & $$-$0.008$ & $$-$1.499$ \\ 
MGwt & $$-$0.006$ & $$-$1.255$ & $$-$0.006$ & $$-$1.341$ \\ 
MGminD & $$-$0.0005$ & $$-$0.179$ & $0.0002$ & $0.096$ \\ 
EPSGwt & $$-$0.001$ & $$-$0.308$ & $$-$0.001$ & $$-$0.248$ \\ 
EPSGminD & $0.0002$ & $0.187$ & $$-$0.0001$ & $$-$0.051$ \\ 
CEOtwt & $0.001$ & $0.156$ & $$-$0.011$ & $$-$1.427$ \\ 
CEOtminD & $$-$0.0001$ & $$-$0.084$ & $$-$0.004$ & $$-$1.006$ \\ 
logRevwt & $0.896$ & $7.028$ & $1.059$ & $11.821$ \\ 
logRevminD & $0.859$ & $3.516$ & $0.994$ & $4.553$ \\ 
logEarnwt & $$-$0.052$ & $$-$0.738$ & $$-$0.064$ & $$-$1.088$ \\ 
logEarnminD & $0.017$ & $0.935$ & $0.020$ & $1.281$ \\ 
logEprofwt & $0.004$ & $0.438$ & $0.001$ & $0.101$ \\ 
logEprofminD & $$-$0.016$ & $$-$1.946$ & $$-$0.008$ & $$-$1.157$ \\ 
logMCapwt & $0.048$ & $0.386$ & $$-$0.010$ & $$-$0.097$ \\ 
logMCapminD & $$-$0.180$ & $$-$0.671$ & $$-$0.178$ & $$-$0.720$ \\ 
logTSRwt & $0.032$ & $2.079$ & $0.012$ & $0.737$ \\ 
logTSRminD & $$-$0.004$ & $$-$0.628$ & $0.001$ & $0.114$ \\ 
\hline \\[-1.8ex] 
\end{tabular} 
\end{table}

\bigskip

It is somewhat surprising that many of these coefficients are significantly negative, but gratifying
 that many are significantly non-zero (or nearly so).  Most also seemed somewhat consistent
 from 2017 to 2018. These results suggest strongly that there is a signal. This is confirmed
 by Table 5, which gives values of the 
mean absolute error (MAE) averaged over the 2 years of prediction.

 \begin{table} \centering 
  \caption{MAE: 2017-8 mean (SD) } 
  \label{MAE} 
\footnotesize 
\begin{tabular}{@{\extracolsep{5pt}} cccccc} 
\\[-1.8ex]\hline 
\hline \\[-1.8ex] 
Method & logRev & logEarn & logEprof & logMCap & logTSR \\ 
\hline \\[-1.8ex] 
cancor & 0.190 (.049) & 1.878 (.271) & 6.050 (.209) & 0.498 (.042) & 7.036 (.192) \\ 
Index & 0.162 (.028) & 1.413 (.020) & 4.940 (.065) & 0.529 (.027) & 6.262 (.162) \\ 
CEOrq & 0.776 (.012) & 1.610 (.015) & 5.081 (.142) & 0.658 (.014) & 6.327 (.170) \\
\hline \\[-1.8ex] 
\end{tabular} 
\end{table} 

\bigskip

\newpage

Clearly, the canonical regression quantile provides better predictions than CEOtot
for all response variables under the more robust MAE criterion. The plots below show
that the moderately large number of negative values for log Economic Profits and Total Shareholder
Return make these variables somewhat anomalous, and this point will be discussed
further below. It was true that the canonical correlation index  outperformed the canonical rq index
 in terms of RMSE (Root Mean
Squared Error), but the difference was not large, and CEOtot was never the better predictor. 
The differences between MAE and RMSE seem mainly a result of the bimodality and outliers.

Finally, some plots for the 2017 predicted responses may clarify the nature of the data. 
Plots for 2018 are quite similar. Figures 1 and 2 present  scatter plots of the 
canonical regression quantile index and CEOtot on the $x$-axis against the canonical regression
quantile $Y$-index ($\alpha ' Y$,  see Figure 1) and against each of the 5 $Y$ variables (Figure 2).
The median regression lines are also plotted.

It is clear that the canonical regression quantile index is predicting its $Y$-index much better than
CEOtot. In fact, if the $Y$-index is fit using both the canonical regression quantile index and CEOtot,
the coefficient of CEOtot is not significantly different from zero, while the coefficient of the index is
highly significant. Regressions of each of the Y-responses on the Index and CEOtot together is 
discussed below.

Considering the plot for logEprof and logTSR, clearly the 
negative values are well-separated from positive ones, and are remarkably large
in absolute value and rather variable. Both
predictors predict the more numerous positive values quite well
(as would be expected for a median predictor), but the CEOtot predictions  seem pulled
down more toward the negative values and thus tend to underestimate the positive ones.
There are only a few
negative responses for other variables, and so these act more like outliers.

Figure 3 gives Q-Q plots of the observed $Y$-response vs. the predicted response from each
of the two methods. The separation of negative and positive responses is clear again, as is the
tendency of CEOtot to overestimate smaller responses and underestimate larger ones.

\begin{figure}[h]
\begin{center}
\includegraphics[height=6in, width=4in ]{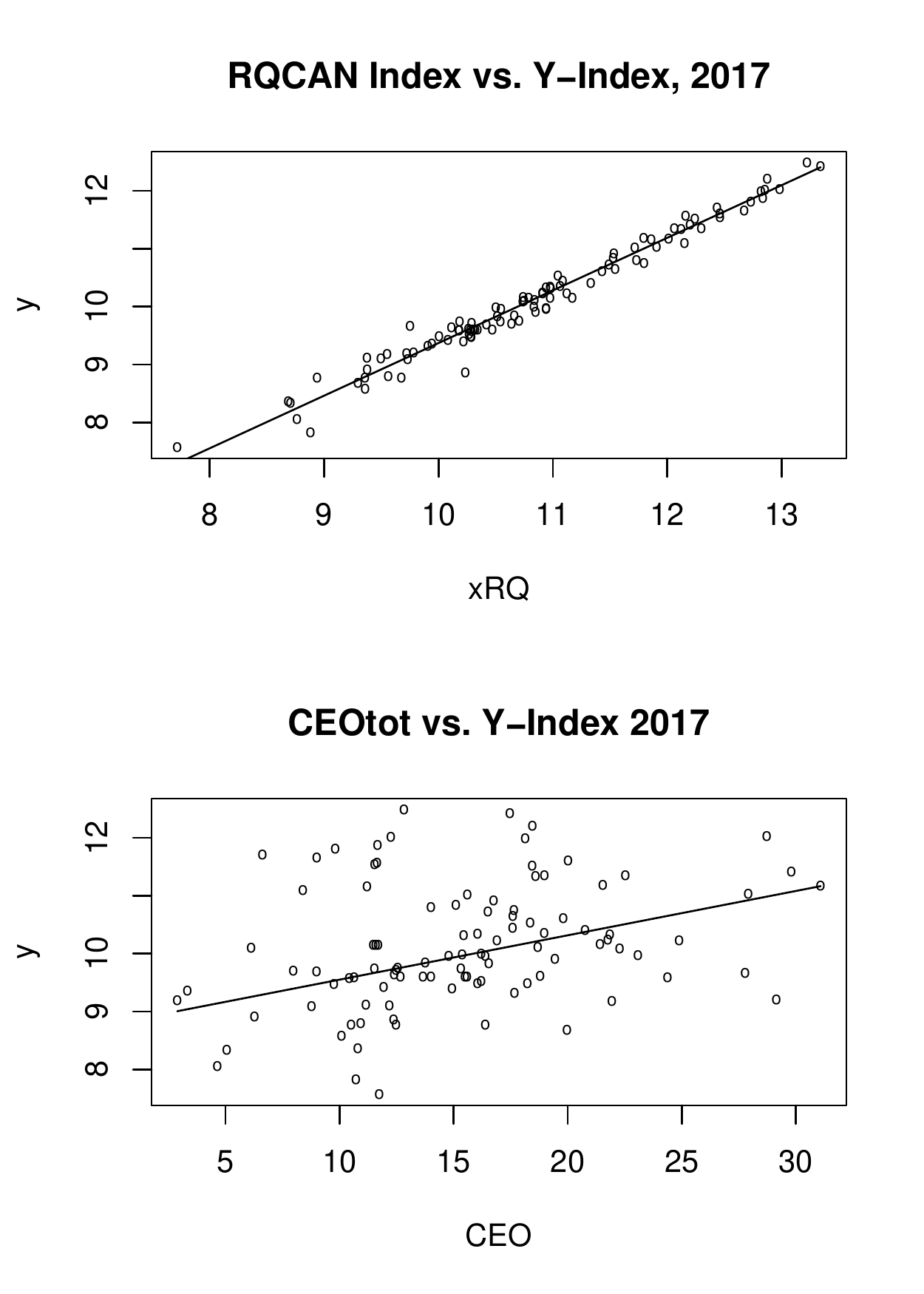}
\caption{2017: response index vs. prediction index and CEOtot}
\end{center}
\par
\label{Yindex}
\end{figure}

\begin{figure}[h]
\begin{center}
\includegraphics[height=7in, width=4in ]{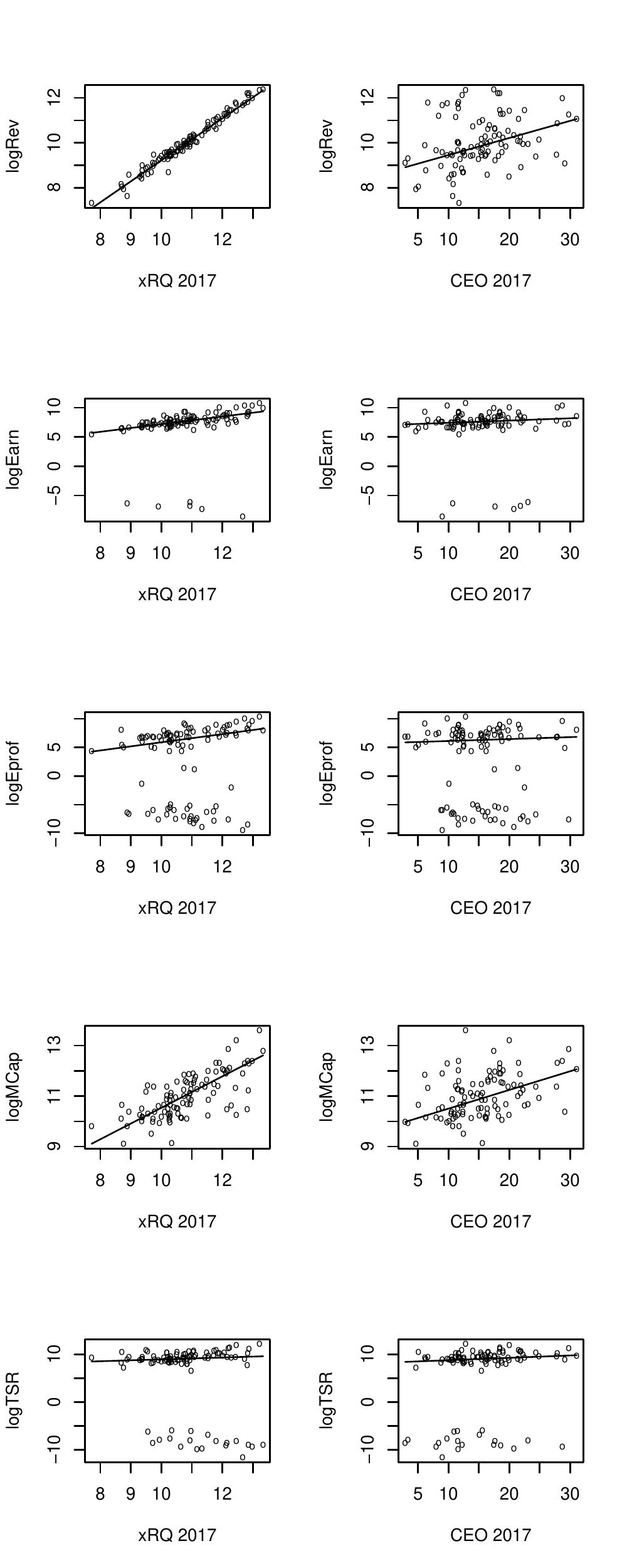}
\caption{2017: Y-variables vs. prediction index and CEOtot}
\end{center}
\par
\label{allY}
\end{figure}

\begin{figure}[h]
\begin{center}
\includegraphics[height=6in, width=4in ]{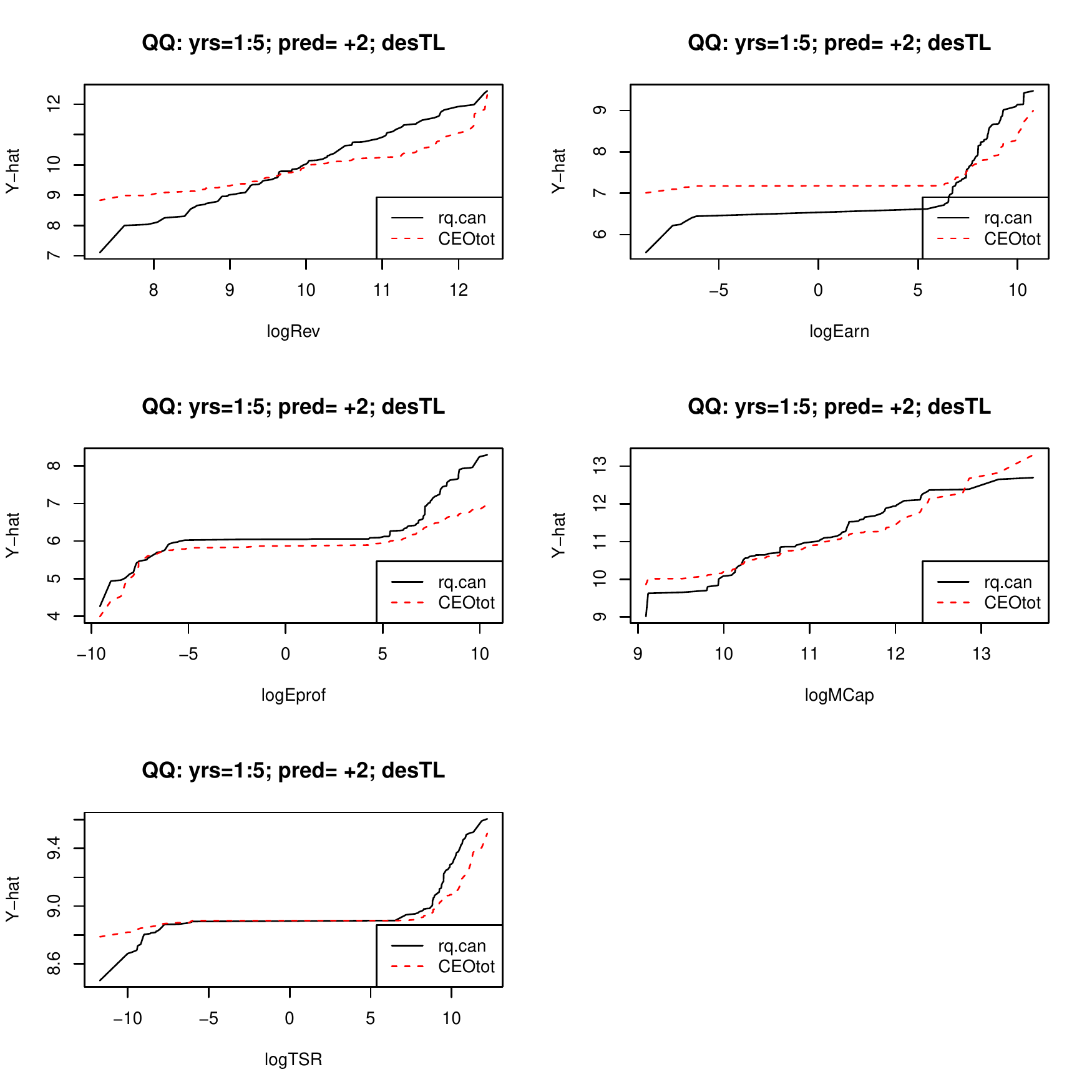}
\caption{2017: Q-Q plot, fit vs. response}
\end{center}
\par
\label{QQ}
\end{figure}

To provide an alternative numerical measure of how much better the index does as compared to CEOtot in
predicting the $Y$-variables, we consider predicting each $Y$ variable from both CEOtot and
the index together (in terms of median regression).  That is, we use the the Index and CEOtot in 2015
to predict 2017 responses, and the Index and CEOtot in 2016 to predict 2018 responses. We use median 
regression with standard errors computed using the Andrews bootstrap (that is, the Indices are recomputed
for each bootstrap replication).  The resulting t-statistics are summarized in Table 6. Note that
the $t$-statistic for the Index tends to be much larger than the $t$ statistic for CEOtot,
 strongly indicating that the index is a much better predictor. In fact, the CEOtot coefficient is negative
 for three responses in 2017.

\begin{table} \centering 
  \caption{t-statistics for CEOtot and Index  } 
  \label{t-stats} 
\begin{tabular}{@{\extracolsep{5pt}} cccccc} 
\\[-1.8ex]\hline 
\hline \\[-1.8ex] 
2017  & logRev & logEarn & logEprof & logMCap & logTSR \\ 
\hline \\[-1.8ex] 
Index15 & $20.33$ & $6.57$ & $2.51$ & $5.35$ & $0.62$ \\ 
CEO15 & $$-$1.49$ & $$-$0.02$ & $$-$1.35$ & $2.35$ & $0.08$ \\ 
\hline
2018 &  &  &  &  &  \\
Index16 & $24.01$ & $3.93$ & $1.46$ & $5.83$ & $0.90$ \\ 
CEO16 & $0.51$ & $1.18$ & $0.11$ & $2.72$ & $1.08$ \\ 
\hline \\[-1.8ex] 
\end{tabular} 
\end{table} 
 
\bigskip

Finally, since a fundamental advantage of regression quantile methods is their ability
to provide information about the entire response distribution (nonparametrically), some runs
were done using the  $\, \tau = .75 \,$ canonical regression quantile to construct indices. As
expected, the .75 canonical regression quantile indices
were not as good as the median ($\tau = .5$) for both
RMSE and MAE. When $\, \rho_{\tau=.75} \,$ was used as a measure of accuracy,
the .75-quantile showed a relatively similar improvement over CEOtot 
except for logEprof and logTSR. This is not surprising in view of the bimodality of this
variable shown in the plots. Note that the $\, \rho_{\tau=.75} \,$ metric weights positive
residuals exactly 3 time more than negative ones, thus exaggerating the cost of underestimation.
The numerical results for MAE and for the .75-quantile loss are given in Tables 7 and 8.

\begin{table} \centering 
  \caption{MAE for .75 quantile} 
  \label{MAE.75} 
\begin{tabular}{@{\extracolsep{5pt}} cccccc} 
\\[-1.8ex]\hline 
\hline \\[-1.8ex] 
 & logRev & logEarn & logEprof & logMCap & logTSR \\ 
\hline \\[-1.8ex] 
cancor & 0.190 (.049) & 1.878 (.271) & 6.050 (.209) & 0.498 (.042) & 7.036 (.192) \\ 
rq.can & 0.192 (.040) & 1.540 (.061) & 5.144 (.064) & 0.625 (.050) & 6.670 (.120) \\ 
CEOrq & 0.972 (.097) & 1.791 (.081) & 5.265 (.104) & 0.827 (.069) & 6.673 (.118) \\ 
CEOls & 0.784 (.018) & 2.012 (.232) & 6.004 (.207) & 0.661 (.011) & 7.069 (.198) \\ 
\hline \\[-1.8ex] 
\end{tabular} 
\end{table} 

\begin{table} \centering 
  \caption{.75 quantile objective function} 
  \label{Mrho.75} 
\begin{tabular}{@{\extracolsep{5pt}} cccccc} 
\\[-1.8ex]\hline 
\hline \\[-1.8ex] 
 & logRev & logEarn & logEprof & logMCap & logTSR \\ 
\hline \\[-1.8ex] 
cancor & 0.095 (.024) & 0.930 (.055) & 3.004 (.042) & 0.248 (.026) & 3.515 (.123) \\ 
rq.can & 0.126 (.028) & 1.104 (.057) & 3.779 (.061) & 0.421 (.045) & 4.878 (.121) \\ 
CEOrq & 0.637 (.092) & 1.248 (.078) & 3.827 (.100) & 0.548 (.068) & 4.873 (.128) \\ 
CEOls & 0.395 (.035) & 1.002 (.054) & 2.985 (.051) & 0.331 (.026) & 3.522 (.114) \\ 
\hline \\[-1.8ex] 
\end{tabular} 
\end{table} 

\bigskip\bigskip\bigskip

{\bf Remarks:} A number of particular specifications were employed in the analysis summarized
above. In fact, many of these specifications were generalized in trials not reported here. Several
runs considered using 4 years of prior data to predict 1, 2, and 3 years ahead. The results
differed in detail, but were qualitatively the same.

Notice that the standard errors for the for logEprof and logTSR are somewhat larger
than other standard errors, reflecting the bimodality of this data. Nonetheless,
the standard error estimates above seem reliable. As noted,
using a weighted delete-$\sqrt{n}$ jackknife gave
similar results, as did the standard bootstrap using total revenue alone to develop the
index. Using a larger resampling size (R = 600 rather than R = 200) also gave
similar standard error estimates, with the same cases of larger standard errors.

The original idea was to use only the $X$-data to develop the indices. Some runs were carried 
out using only the X-data (all the $X$-data and not just the 2 summary aggregates)
from 2 to 4 earlier years. This approach did not produce very good predictors.
Some attempts at variable selection were also tried. In almost all cases, the variable
selection methods suggested that all variables were needed, and so differed very little (or
not at all) from the results presented here.

\section{Conclusions.}

The Canonical Regression Quantile method has been developed in analogy with Canonical Correlations.
Basically, we seek linear combinations of explanatory and response variables with the $L_2$ normalization 
of Canonical Correlations replaced by an $L_1$ normalization and with mean squared error replaced
by the regression quantile objective function. We have focussed on only the most predictive linear
combinations (analogous to the leading canonical correlation). It is conceptually clear that canonical
regression quantiles can be extended by seeking the most predictive indices subject to the constraint 
that they be orthogonal to the leading indices. Since the condition for orthogonality is linear in the
new linear coefficients, the computations are essentially the same application of constrained regression
quantiles as for the leading indices. The process could clearly be continued to provide a dimension reduction
method entirely analogous to Canonical Correlations. The main conceptual problem is the imposition of
orthogonality: outside normal models, the resulting indices will not be independent, and it seems clear
that specific applications may suggest alternative conditions, both for canonical quantile regression as 
well as for traditional canonical correlations. Some preliminary explorations are promising. 

The application to financial data show that Canonical Regression Quantiles can provide useful
information for the scientist. Additional analysis and financial development in 
Haimberg and Portnoy (2021) indicate that, at least when economic conditions are stable, there is
information available that might provide better ways to set CEO compensation so as to improve 
company performance. 


\end{document}